\documentclass[aps,nofootinbib,prc,twocolumn,preprintnumbers, floatfix, superscriptaddress]{revtex4-1}

 \pdfoutput=1
\usepackage[usenames,dvipsnames]{color}  
\usepackage{graphicx}
\usepackage[utf8]{inputenc}
\usepackage{caption}
\usepackage{subcaption}
\usepackage{diagbox}

\usepackage{amsmath}
\usepackage{amssymb}
\usepackage[colorlinks=true,citecolor=darkred,urlcolor=darkred, pdfborder={0 0 0}]{hyperref}
\usepackage[normalem]{ulem}

\usepackage[T1]{fontenc}
\usepackage{array}
\usepackage{booktabs}
\usepackage{mathrsfs}
\usepackage{multirow}
\usepackage{tabularx}
\usepackage[utf8]{inputenc}
\usepackage[normalem]{ulem}
\definecolor{darkred}{rgb}{0.6,0,0}

\definecolor{linkcolor}{rgb}{0,0,0.5}

\def\gsim{\raise0.3ex\hbox{$\;>$\kern-0.75em\raise-1.1ex\hbox{$\sim\;$}}}
\def\lsim{\raise0.3ex\hbox{$\;<$\kern-0.75em\raise-1.1ex\hbox{$\sim\;$}}}

\def\beqn#1{\begin{equation}\label{#1}}
\def\eeqn{\end{equation}}

\def\beqa#1{\begin{eqnarray}\label{#1}}
\def\eeqa{\end{eqnarray}}

\def\Z2{$\mathcal{Z_2}$}

\newcommand {\ignore}[1]{}

\def\cevns{CE$\nu$NS~}

\def\321{$\mathrm{SU(3) \otimes SU(2) \otimes U(1)}$ }

\def\be{\begin{equation}}

\def\ee{\end{equation}}
\def\barr{\begin{array}}
\def\earr{\end{array}}

\def\nn8{\nonumber\\[2pt]}
\def\l{\left}
\def\r{\right}
\def\dis{\displaystyle}
\def\ed{\end{document}}

\def\cevns{CE$\nu$NS~}

\def\be{\begin{equation}}

\def\ee{\end{equation}}
\def\barr{\begin{array}}
\def\earr{\end{array}}

\def\l{\left}
\def\r{\right}
\def\dis{\displaystyle}
\def\ed{\end{document}}

\def\f{\frac}

\def\d{{\text{d}}}

\begin{document}

\title{Coherent elastic neutrino-nucleus scattering (CE$\nu$NS) event rates for Ge, Zn and Si detector materials}

\author{T. S. Kosmas}\email{hkosmas@uoi.gr}\affiliation{Division of Theoretical Physics, University of  Ioannina, GR 45110 Ioannina, Greece}

\author{V. K. B. Kota}\email{vkbkota@prl.res.in}\affiliation{Physical
Research Laboratory, Ahmedabad 380 009, India}

\author{D. K. Papoulias}\email{d.papoulias@uoi.gr} \affiliation{Division of Theoretical Physics, University of  Ioannina, GR 45110 Ioannina, Greece}

\author{R. Sahu}\email{rankasahu@gmail.com}\affiliation{National Institute
of Science and Technology, Palur Hills, Berhampur 761 008, Odisha, India}

\begin{abstract}
Realistic nuclear structure calculations are presented for the event rates due to coherent elastic neutrino-nucleus scattering (CE$\nu$NS), assuming neutrinos from pion decay at rest, from nuclear reactors, and from Earth's interior. We focus on the currently interesting germanium isotopes, $^{70,73,76}$Ge, which constitute detector materials of the recently planned CE$\nu$NS experiments. We study in addition the potential use of $^{64,70}$Zn and $^{28}$Si isotopes as promising CE$\nu$NS detectors. From nuclear physics perspectives, recently, calculations have been carried out within the framework of the deformed shell model (DSM), based on realistic nuclear forces, and assessed on the reproducibility of spectroscopic nuclear properties. The high confidence level acquired by their agreement with experimental results and by their comparison with other mostly phenomenological calculations encouraged the use of DSM to extract predictions for the CE$\nu$NS event rates of the above isotopes. Our detailed estimation of the nuclear physics aspects of the recently observed neutral current coherent neutrino-nucleus scattering may shed light on unravelling the still remaining uncertainties for the
CE$\nu$NS process within and beyond the standard model.
\end{abstract}

\maketitle

\section{Introduction}

More than four decades ago, Freedman~\cite{Freedman:1973yd} proposed the measurement of the neutral current coherent 
elastic neutrino-nucleus scattering (CE$\nu$NS) when low-energy neutrinos scatter off nuclei. This process, however, 
was observed for the first time very recently by the COHERENT collaboration~\cite{Akimov:2017ade} using the sodium-doped CsI detector at the Spallation Neutron Source (SNS) at Oak Ridge National Laboratory. The process was, 
subsequently, observed at the SNS using also a liquid argon (LAr) detector~\cite{Akimov:2020pdx}. 
The observation of \cevns has opened up new opportunities to test the
predictions of the standard model (SM)~\cite{Cadeddu:2017etk,Papoulias:2019lfi,Sahu:2020kwh,Tomalak:2020zfh,Co:2020gwl}, while a  precise
measurement of this process may offer a way to constrain the particle physics parameters of theories beyond the SM 
\cite{Papoulias:2017qdn} (the recent constraints extracted from \cevns are summarized in Ref.~\cite{Papoulias:2019xaw}).

The detection signal of CE$\nu$NS, i.e., the low-energy recoil of the target nucleus, is an experimental challenge 
while the uncertainties associated with the relevant measurements should be minimized and the accuracy of CE$\nu$NS 
experimental method must be improved. Toward this purpose, many planned experiments for measuring CE$\nu$NS are based 
on the well known germanium detectors~\cite{Yuri-Efremenko}, while zinc and silicon are also promising detector materials 
for neutrino-nucleus cross section measurements~\cite{Tsakstara:2012yd}. 
Such ongoing and designed detectors are CONUS~\cite{CONUS:2020skt}, 
$\nu$GEN~\cite{Belov:2015ufh}, TEXONO~\cite{TEXONO:2020vnv}, COHERENT~\cite{COHERENT:2018gft},
RICOCHET~\cite{Billard:2016giu}, MINER~\cite{MINER:2016igy}, NUCLEUS~\cite{NUCLEUS:2019igx}, CONNIE~\cite{CONNIE:2019xid}, 
Coherent Captain-Mills (CCM)~\cite{CCM}, European Spallation Source
(ESS)~\cite{Baxter:2019mcx}, vIOLETA~\cite{Fernandez-Moroni:2020yyl} and SBC~\cite{SBC:2021yal}
experiments. The employment of pure Ge detectors in measuring rare event processes has shown appreciably
good sensitivity, while in CE$\nu$NS some combinations of detection media have been chosen 
and proposed to be utilized due to other experimental criteria~\cite{Galindo-Uribarri:2020huw}. 
Toward the latter purposes, Zn and Si isotopes may offer advantageous
combinations to reduce the systematic errors of CE$\nu$NS experiments instead of a single element
\cite{Yuri-Efremenko}.

Theoretically, it was known that roughly speaking the CE$\nu$NS cross section has a quadratic dependence 
on the neutron number of the target nucleus ($\propto N^2$) which is attributed to the different strength
of the respective couplings with which the protons and the neutrons of the atomic nuclei interact with 
the intermediate $Z_0$ boson; see, e.g., Ref.~\cite{Drukier:1984vhf}.
The ground-state to ground-state transition channel, which is possible in neutral current neutrino-nucleus
scattering, appears enhanced due to the fact that the proton and neutron amplitude phases corresponding to 
a neutrino scattering off nucleons are added coherently~\cite{Lindner:2016wff}, and dominates the process 
at low energies. On the other hand, the incoherent scattering cross sections are much smaller and demonstrate 
some well pronounced peaks of specific multipole excitations~\cite{Pirinen:2018gsd}. Such detailed calculations have been performed 
previously for various nuclear isotopes (see, e.g., Refs.~\cite{Tsakstara:2011zzc, Tsakstara:2011zzd}).  

With respect to physics beyond the SM, nonstandard interactions (NSIs) is a widely used formalism that can phenomenologically describe a large family of new physics interactions, in particular those involving novel vector or axial vector processes~\cite{Flores:2020lji}. 
Constraints on NSIs exist from the analysis of available \cevns data by COHERENT with CsI~\cite{Liao:2017uzy,Papoulias:2017qdn,Giunti:2019xpr} and LAr~\cite{Miranda:2020tif}. Moreover, extensions of the NSI formalism, namely neutrino generalized interactions (NGIs) can accommodate scalar, pseudoscalar and tensor interactions~\cite{AristizabalSierra:2018eqm}.   

Over the years, the deformed shell model (DSM) based on Hartree-Fock states with angular momentum projection 
and band mixing has been found to be quite successful in describing several nuclear properties like spectroscopic 
properties including spectroscopy of $N = Z$ odd-odd nuclei with isospin projection~\cite{Srivastava:2017elr}, 
the coherent and incoherent neutral current $\mu \rightarrow e$ conversion in the field of nuclei~\cite{Kosmas:2003xr} and double-$\beta$ decay half-lives~\cite{Sahu:2013yna, Sahu:2014nga}. Recently, we have calculated event rates 
for weakly interacting massive particle (WIMP) scattering off $^{73}$Ge \cite{Sahu:2017czz} and elastic and 
inelastic scattering of neutrinos and WIMPs on nuclei~\cite{Papoulias:2018uzy, Sahu:2020kwh}.  
Our aim in this work is to provide reliable theoretical predictions for event rates of neutrino-nucleus scattering 
involving $^{70,73,76}$Ge, $^{64,70}$Zn and $^{28}$Si isotopes by using DSM for the nuclear structure functions 
needed for the event rate calculations. Motivated by the various experimental facilities, first, we focus on 
pion decay at rest ($\pi$-DAR) and reactor antineutrino sources. We furthermore consider geoneutrinos, which are 
expected to contribute sizable the overall neutrino background signal at the next-generation large-scale detectors 
planned to look for light WIMPs~\cite{Gelmini:2018gqa}. 

Finally, we quantify the percentage differences on the expected number of events obtained with the DSM, as compared 
to those relying on the widely used approximate form factor parametrizations, at different energy regimes, as well as
to those obtained with different nuclear physics methods~\cite{VanDessel:2020epd}. We stress that, in extracting the 
percentage differences we rely on the number of events of the \cevns which is the most relevant experimental observable. 

The paper has been organized as follows: In Sec.~\ref{sec:formalism}, we discuss the \cevns formalism and our adopted nuclear structure method calculated in the framework of the DSM. Then, in Sec.~\ref{sec:event-rates}, we present the theoretical event rates and we discuss the level of inconsistency with respect to similar calculations involving effective nuclear form factors. Finally, our conclusions are summarized in Sec.~\ref{sec:conclusions}.

\section{basic formalism}
\label{sec:formalism}

In this section, we present the basic formalism for calculating the CE$\nu$NS event rates at different facilities aiming to detect signals induced by $\pi$-DAR, reactor or geoneutrinos. We pay special attention to discussing the key ingredients of the nuclear physics aspects that have been taken into consideration. In particular, the nuclear form factors are obtained using nuclear wavefunctions determined by DSM, 
of which the confidence level is well-established through the reproducibility of spectroscopic results for the nuclear isotopes of interest (see below).

Before embarking to the nuclear physics calculations, it is worth mentioning that, interesting studies of physics beyond the standard model include deviations from unitarity~\cite{Miranda:2020syh}, sterile neutrinos~\cite{Kosmas:2017zbh, Canas:2017umu, Miranda:2020syh}, 
neutrino magnetic moments~\cite{Wong:2005vg, Miranda:2019wdy,Papoulias:2019txv}, nonstandard interactions~\cite{AristizabalSierra:2018eqm,Giunti:2019xpr, Miranda:2020zji,Denton:2020hop}, light new physics~\cite{Abdullah:2018ykz,Flores:2020lji}, dark matter~\cite{Dutta:2019nbn,delaVega:2021wpx}, etc.

\subsection{CE$\nu$NS differential cross section}

Neutrinos with energies below some tens of MeV
predominately conserve the integrity of nucleons in 
neutrino-quark interactions with $Z_0$-boson exchange, allowing
us to consider the \cevns process using an effective neutrino nucleon interaction in which the nucleon
current is a sum of vector and axial currents. The differential \cevns cross section with respect to 
the nuclear recoil energy $T_A$ (the axial vector contributions is neglected in this work) reads~\cite{Papoulias:2017qdn}
\begin{equation}
\frac{\d \sigma}{\d T_A} = \frac{G_\text{F}^2 m_A}{2 \pi} \mathcal{Q}_W^2 \left(2 - \frac{m_A T_A}{E_\nu^2} \right)  \, ,
\label{eq:cevns-xsec}
\end{equation}
where $G_\text{F}$ is the Fermi's constant, $E_\nu$ is the incoming neutrino energy, while $Z$ and $N=A-Z$  denote the number of protons and neutrons, respectively.  The vector weak charge, $\mathcal{Q}_W$, encapsulates the information from the nuclear structure and is written in terms of the proton and neutron form factors $F_{p,n}(q^2)$ as
\begin{equation}
\mathcal{Q}_W = g^V_p Z F_p(q^2) + g^V_n N F_n(q^2) \, ,
\label{eq:Qw}
\end{equation}
where the proton and neutron couplings are expressed as $g^V_p = 1/2 - 2 \sin^2 \theta_W$ and $g^V_n = -1/2$, respectively, and $q=\dis\sqrt{2m_AT_A}$ denotes the magnitude of the 3-momentum transfer.

For the low energies involved in CE$\nu$NS, our calculations consider the low-energy limit of the weak mixing angle running, and hence we assume $\sin^2\theta_W = 0.2381$~\cite{Kumar:2013yoa}. Note, that due to the smallness of the proton coupling, the \cevns cross section scales with a characteristic $N^2$ dependence. Finally, the nuclear mass is calculated as $m_A=Z m_p + N m_n -B$, where the nuclear binding energy $B$ is taken from Ref.~\cite{Wang:2017} (the nucleon masses are taken to be $m_p = 938.28$ MeV and $m_n = 939.57$ MeV). 

\subsection{CE$\nu$NS event rates}
 
The differential and integrated event rates of \cevns, after defining all the parameters in Eq.(\ref{eq:cevns-xsec}), are calculated by
\be
\frac{\d R}{d T_A} = \mathcal{K} \dis\int_{E_{\nu,\text{min}}}^{E_{\nu,\text{max}}} \l[\dis\frac{\d \sigma}{\d T_A} \l(E_
\nu, T_A\r)\r]\;\; \lambda_{\nu} (E_\nu)\, d E_\nu
\label{eq.6}
\ee
where $\lambda_\nu (E_\nu)$ represents the relevant neutrino energy distribution function characterizing the specific neutrino source.  
The normalization factor $\mathcal{K}$ is given by $\mathcal{K}= t_\text{run} \Phi_\nu N_\text{targ} $, with $N_\text{targ} = \dis\frac{m_\text{det}\,N_A}{M_r}$. 
Here, $t_\text{run}$ is taken as $1\;$yr, $N_\text{targ}$ is the number of target nuclei and $\Phi_\nu$ is the neutrino flux  normalization. In the 
calculation of $N_\text{targ}$, the detector mass $m_\text{det}$ is assumed to be 1 kg for $\pi$-DAR and reactor neutrinos and 1 ton for geoneutrinos. 
Similarly, $M_r$ is the molar mass (atomic weight) and $N_A$ is the Avogadro number ($N_A=6.022 \times 10^{23}$)~\cite{Cadeddu:2017etk}. 

For the case of $\pi$-DAR neutrinos, we consider the specifications of the SNS at Oak Ridge with $r=0.08$ being the number of emitted 
neutrinos per flavor for each proton on target (POT) and $N_\text{POT} \approx 2.1 \times 10^{23}$ denoting the number of protons on target per 
year~\cite{Akimov:2017ade}. The SNS flux is then obtained as $\Phi_\nu = \dis\frac{r \cdot N_\text{POT}}{4\pi L^2}$, which for a typical detector 
baseline of $L \approx 20$ m, evaluates to $\Phi_\nu \approx 1 \times 10^7~\mathrm{s^{-1}~cm^{-2}}$.
Finally, the neutrino energy distribution functions $\lambda_\nu (E_\nu)$ for 
SNS neutrinos are~\cite{Louis:2009zza},
\be
	\lambda_\nu (E_\nu)= \left\{   
	\begin{array}{llc}
	&     \delta \left(E_\nu - \frac{m^2_\pi - m^2_\mu} {2 m_\pi}\right) 
		\;\;\;\;&\text{prompt} \quad \nu_\mu \, ,\\
	&     \frac{64 E^2_\nu}{m^3_\mu}
		\left(\frac{3}{4} -\frac{E_\nu}{m_\mu}\right)\;\;\;\;&\text{delayed} \quad \nu_e \, ,\\
		&\frac{192 E^2_\nu}{m^3_\mu} \, ,
		\left(\frac{1}{2}-\frac{E_\nu}{m_\mu}\right)\;\;\;\;&\text{delayed} \quad \bar{\nu}_\mu \, .
\end{array}\right.
\ee

In the present work, the reactor antineutrino energy-distribution which assumes the fission products of 
$^{235}$U, $^{238}$U, $^{239}$Pu and $^{241}$Pu is taken from  Ref.~\cite{Mention:2011rk}. We note that,
due to the lack of experimental data for $E_\nu< 2 ~\text{MeV}$, the theoretical calculation of 
Ref.~\cite{Kopeikin:1997ve} is employed, while we assume a typical flux of 
$\Phi_\nu \approx 1 \times 10^{13}~\mathrm{s^{-1} \,  cm^{-2}}$. Similarly, for the case of geoneutrinos 
the corresponding antineutrino energy distributions for the K, Th and U chains are taken from Ref.~\cite{huang:2013}. 
It is worth noting that presently the flux uncertainties are quite large due to low statistics~\cite{Ludhova:2013hna} 
while the geoneutrino flux depends largely on the location~\cite{Gelmini:2018gqa}. Here, we comply with the 
normalizations quoted in Ref.~\cite{OHare:2020lva} which correspond to the location of the  Gran Sasso National Laboratory (LNGS).

The differential number of events is obtained through the convolution of the differential cross section with the neutrino-energy distribution  $\lambda_\nu (E_\nu)$.  The lower integration limit in Eq.(\ref{eq.6}) is trivially obtained from the \cevns kinematics and reads
\begin{equation}
E_{\nu,\text{min}} = \frac{1}{2} \left(T_A  + \sqrt{T_A^2 + 2 m_A T_A} \right) \approx \dis\sqrt{\f{m_AT_A}{2}} \, .
\end{equation}
Note that $E_{\nu,\text{max}}= m_\mu/2 = 52.8$ MeV for the case of $\pi$-DAR neutrinos, while $E_{\nu,\text{max}} \approx 9.5$~MeV 
for reactor neutrinos and $E_{\nu,\text{max}}\approx(1.3,~2.3, 4.5)$~MeV for the (K, Th, U) geoneutrinos, respectively.

Finally, an additional integration over the nuclear recoil energy $T_A$, from a threshold energy $T_A^\text{thres}$ up to 
a maximum energy $T_A^\text{max}=\frac{2 E_{\nu,\text{max}}^2}{2 E_{\nu,\text{max}} + m_A }\approx\frac{2 E_{\nu,\text{max}}^2}{m_A}$,  
needs to be performed in order to obtain the expected number of events.

\subsection{Deformed shell model}

The details of the deformed shell model have been described in our earlier publications 
(for details see Ref.~\cite{ks-book}).  In this model, for a given
nucleus, starting with a model space consisting of a given set of spherical single
particle (sp) orbitals with single-particle energies (spe) and an effective two-body interaction specified 
by its two-body matrix elements (TBME), the
lowest energy intrinsic states are obtained by solving the Hartree-Fock (HF)
single-particle equation self-consistently. We assume axial symmetry, while  excited
intrinsic configurations are obtained by making particle-hole excitations over
the lowest intrinsic state.  Since the intrinsic 
states denoted by $\chi_K(\eta)$ do not have definite angular momenta, states of good  angular momentum 
are projected from the latter
which can be  written in the form
\begin{equation}
| \psi^J_{MK}(\eta) \rangle = \frac{2J+1}{8\pi^2\sqrt{N_{JK}}}\int d\Omega D^{J^*}_{MK}(\Omega)R(\Omega)| \chi_K(\eta) \rangle \, ,
\label{eqn.24}
\end{equation}
where $N_{JK}$ is the normalization constant given by
\begin{equation}
N_{JK} = \frac{2J+1}{2} \int^\pi_0 d\beta \sin \beta d^J_{KK}(\beta)\langle \chi_K(\eta)|e^{-i\beta J_y}|\chi_K(\eta) \rangle  \, .
\label{eqn.25}
\end{equation}
In Eq.(\ref{eqn.24}), $\Omega$ represents the Euler angles ($\alpha$, $\beta$,
$\gamma$), and $R(\Omega)=\exp(-i \alpha  J_z) \exp(-i \beta J_y) \exp(-i \gamma J_z)$ 
represents the general rotation operator.  However, it is worth noting that the good
angular momentum states, projected from different intrinsic states, are not in
general orthogonal to each other.  Hence they are orthonormalized and then 
band mixing calculations are performed.  
The resulting eigenfunctions are of the form
\begin{equation}
\vert\Phi^J_M(\eta) \rangle \, =\,\sum_{K,\alpha} S^J_{K \eta}(\alpha)\vert 
\psi^J_{M K}(\alpha)\rangle \, ,
\label{phijm}
\end{equation}
with  $S^J_{K \eta}(\alpha)$ being the expansion coefficients.
The nuclear matrix elements occurring in the calculation of event rates
are evaluated using the wave functions 
$| \Phi^J_M(\eta) \rangle$. We finally stress that the DSM is well established enough to be a 
successful model for transitional nuclei with $A$=60--90
~\cite{ks-book}, while recently, it has also been used successfully for heavier nuclei like 
$^{127}$I, $^{133}$Cs and $^{133}$Xe~\cite{Sahu:2020kwh}.

\section{Results for CE$\nu$NS event rates}
\label{sec:event-rates}

The nuclei selected in the present study, $^{70,73,76}$Ge, $^{64,70}$Zn and $^{28}$Si, are of current experimental 
interest and therefore calculations for CE$\nu$NS event rates that take into account the details of the 
nuclear structure are important. Firstly, the Ge isotopes are discussed in 
Ref.~\cite{Galindo-Uribarri:2020huw} with the most relevant experiments being the COHERENT~\cite{Akimov:2017ade,Akimov:2020pdx}, 
the CONUS~\cite{Bonet:2020ntx,Lindner:2016wff}, $\nu$GEN~\cite{Belov:2015ufh} and the TEXONO~\cite{Wong:2005vg} which 
use ionization-based Ge-semiconductors. Similarly, the (Si, Zn, Ge) isotopes have been chosen as the detector materials 
of the MINER~\cite{MINER:2016igy}, NUCLEUS~\cite{NUCLEUS:2019igx}, and RICOCHET~\cite{Billard:2016giu}, which employ 
cryogenic detectors.

We should, moreover, add that high-purity germanium (HPGe) detectors are used, for example, in CONUS experiment where 
the detector is located at $17.1$ m from the reactor core (4 detectors each with $\approx 1$ kg) and the expected $\bar{\nu}_e$ 
flux is $2.3 \times 10^{13}~\mathrm{s^{-1}cm^{-2}}$; see e.g., Ref.~\cite{Bonet:2020ntx}. Also, for the Chooz experiment 
Ge-based and metallic Zn-based detectors of mass $m_\text{det} \approx 10$ kg are under deployment, with a reported threshold 
as low as 100 eV~\cite{Billard:2016giu}.

\begin{figure*}
\includegraphics[width=0.4\linewidth]{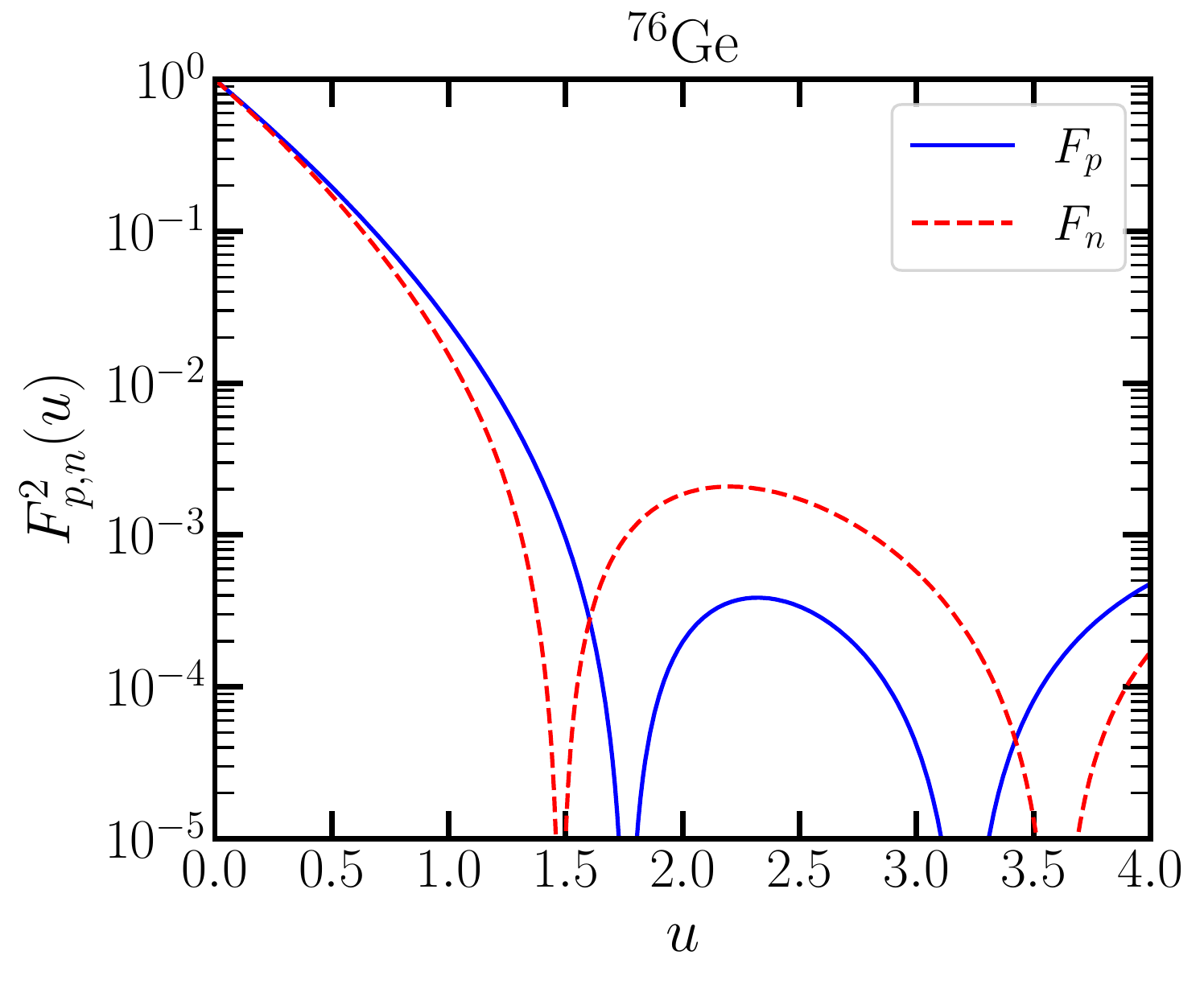}
\includegraphics[width=0.4\linewidth]{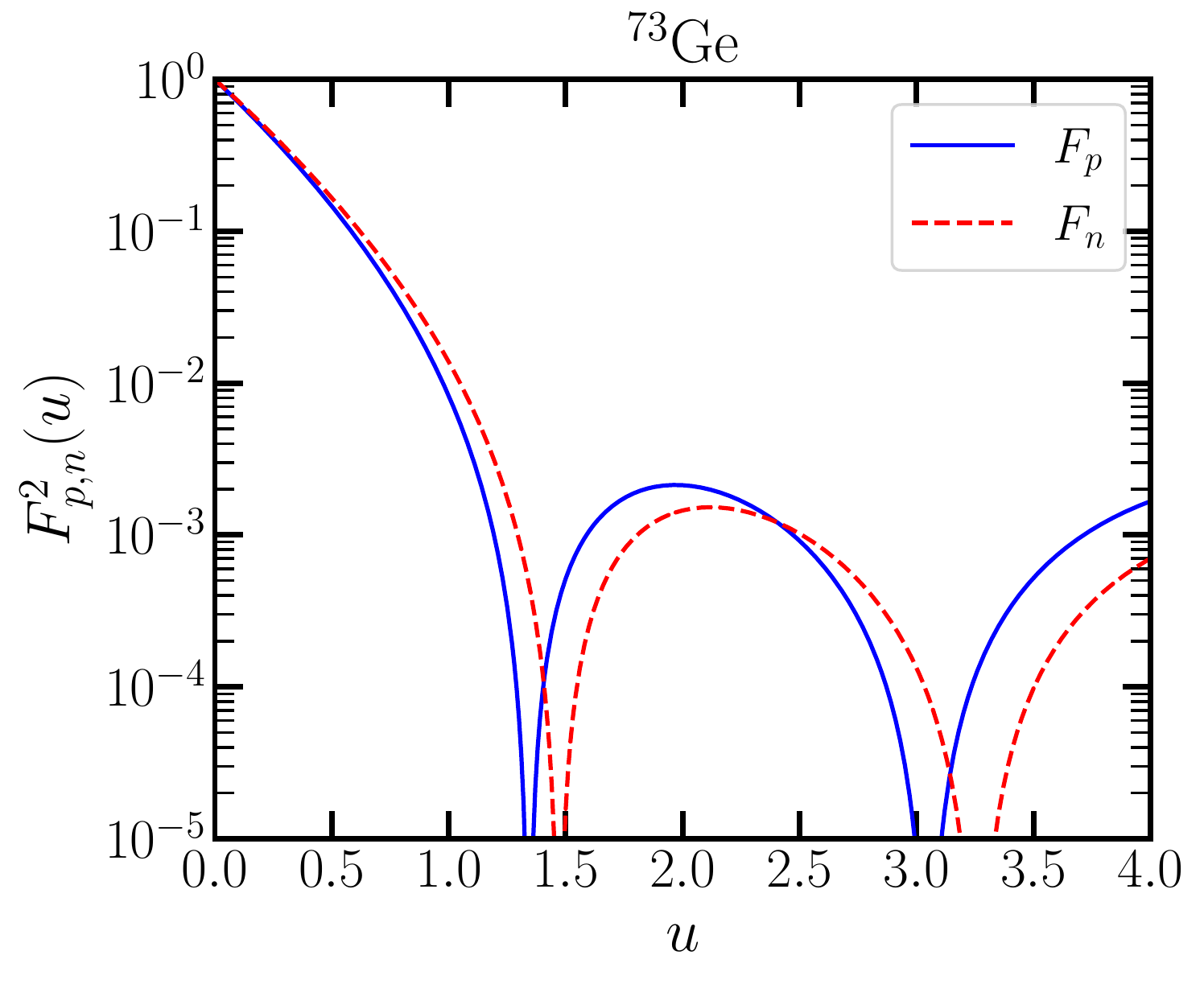}
\includegraphics[width=0.4\linewidth]{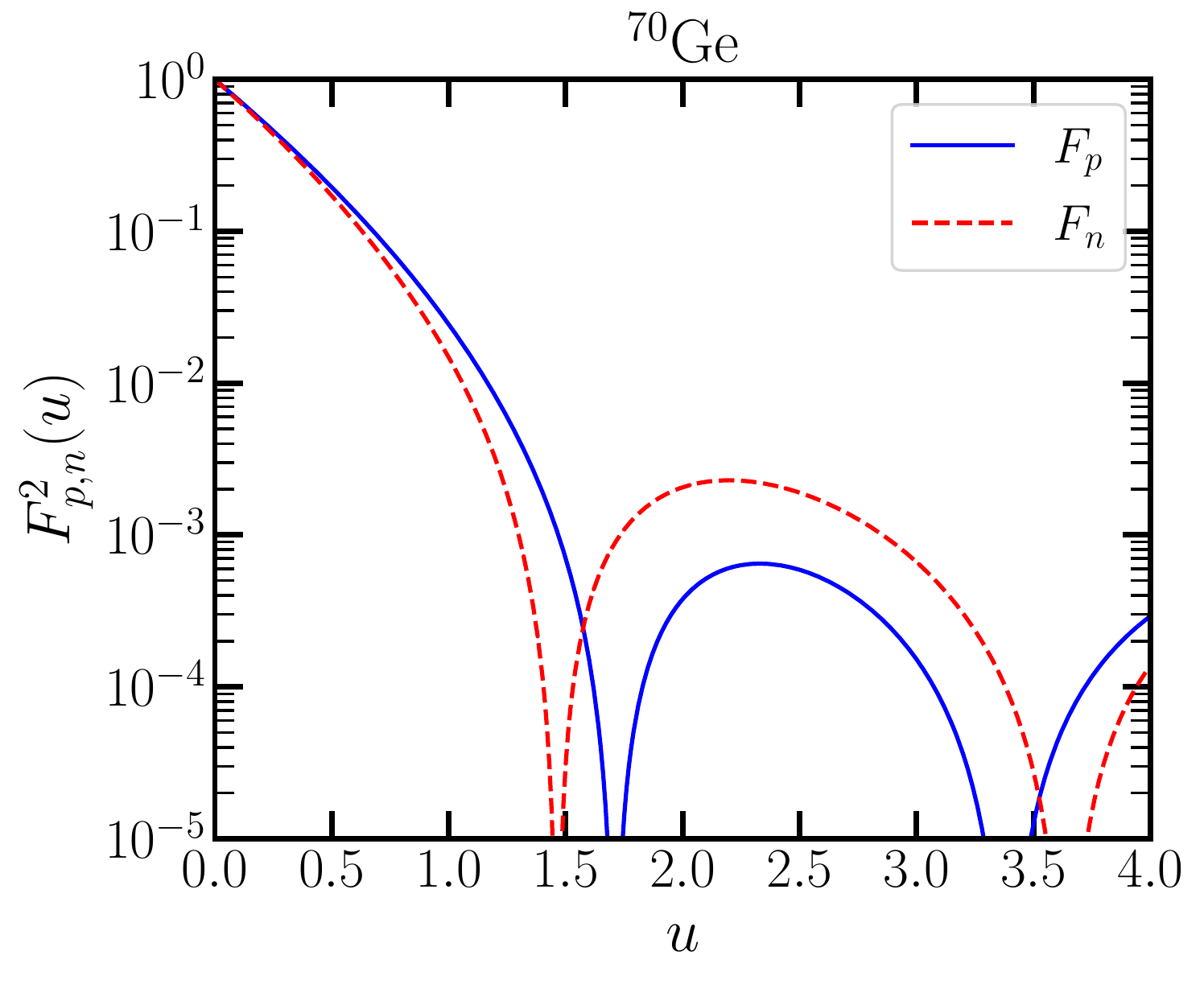}
\includegraphics[width=0.4\linewidth]{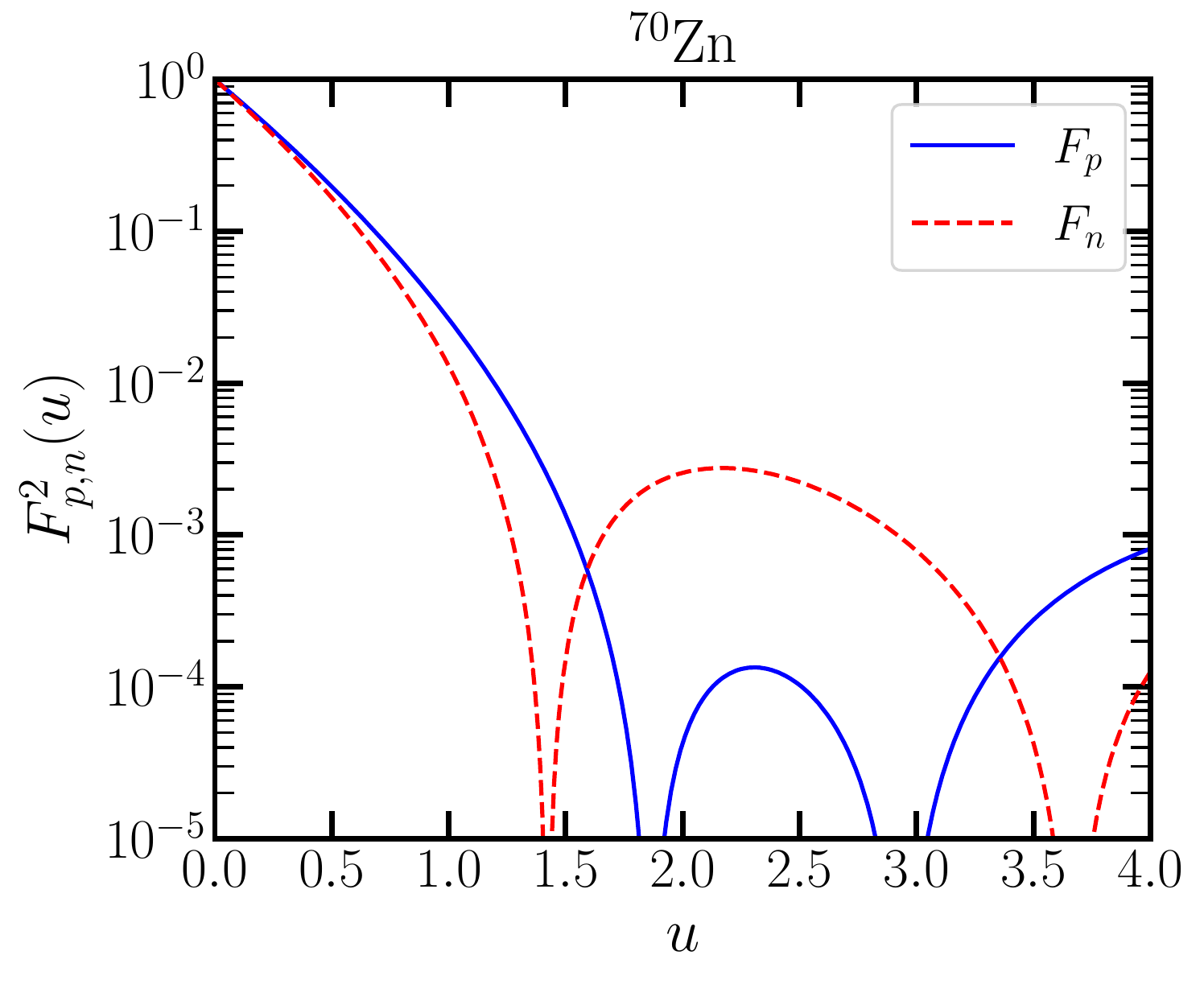}
\includegraphics[width=0.4\linewidth]{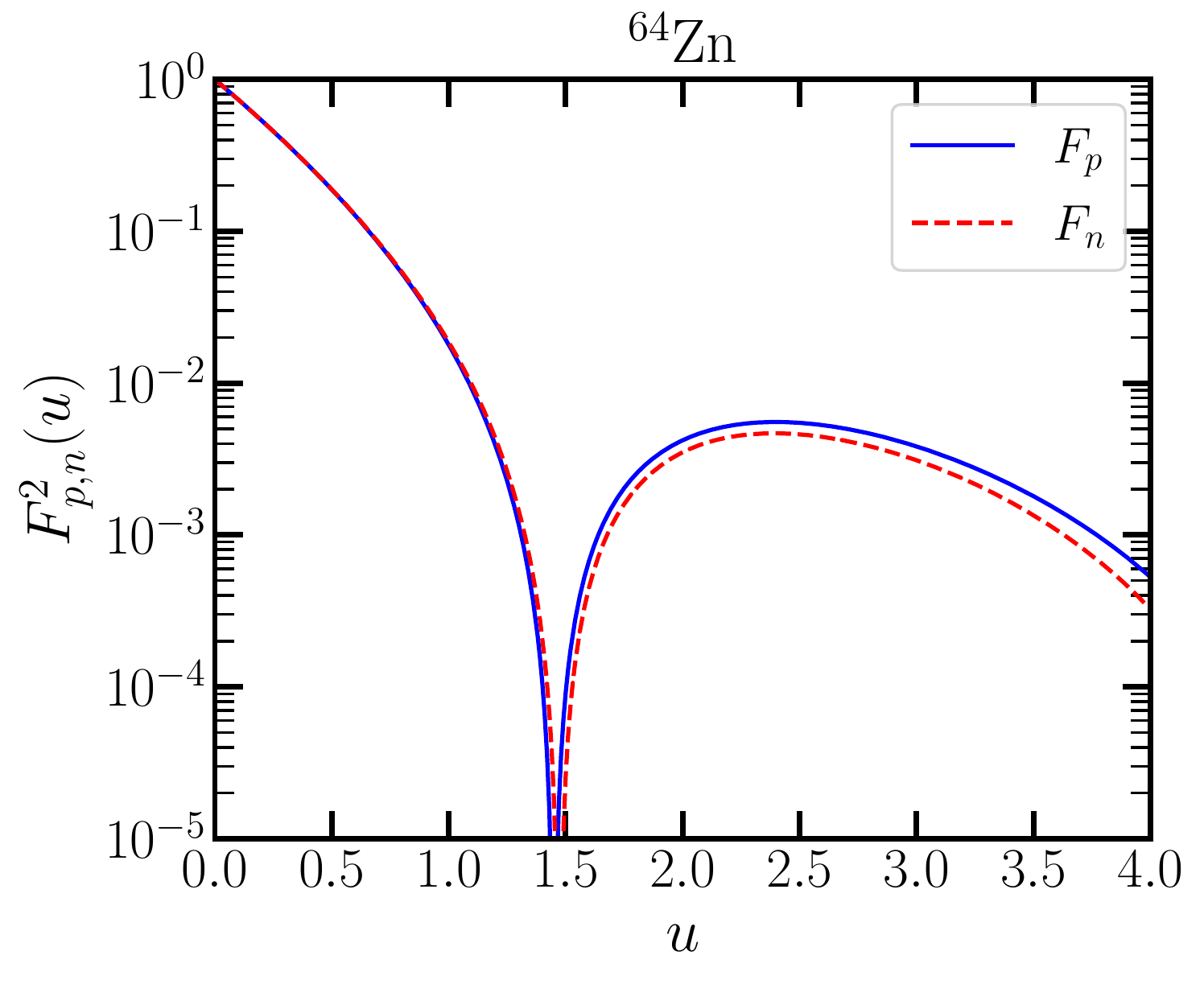}
\includegraphics[width=0.4\linewidth]{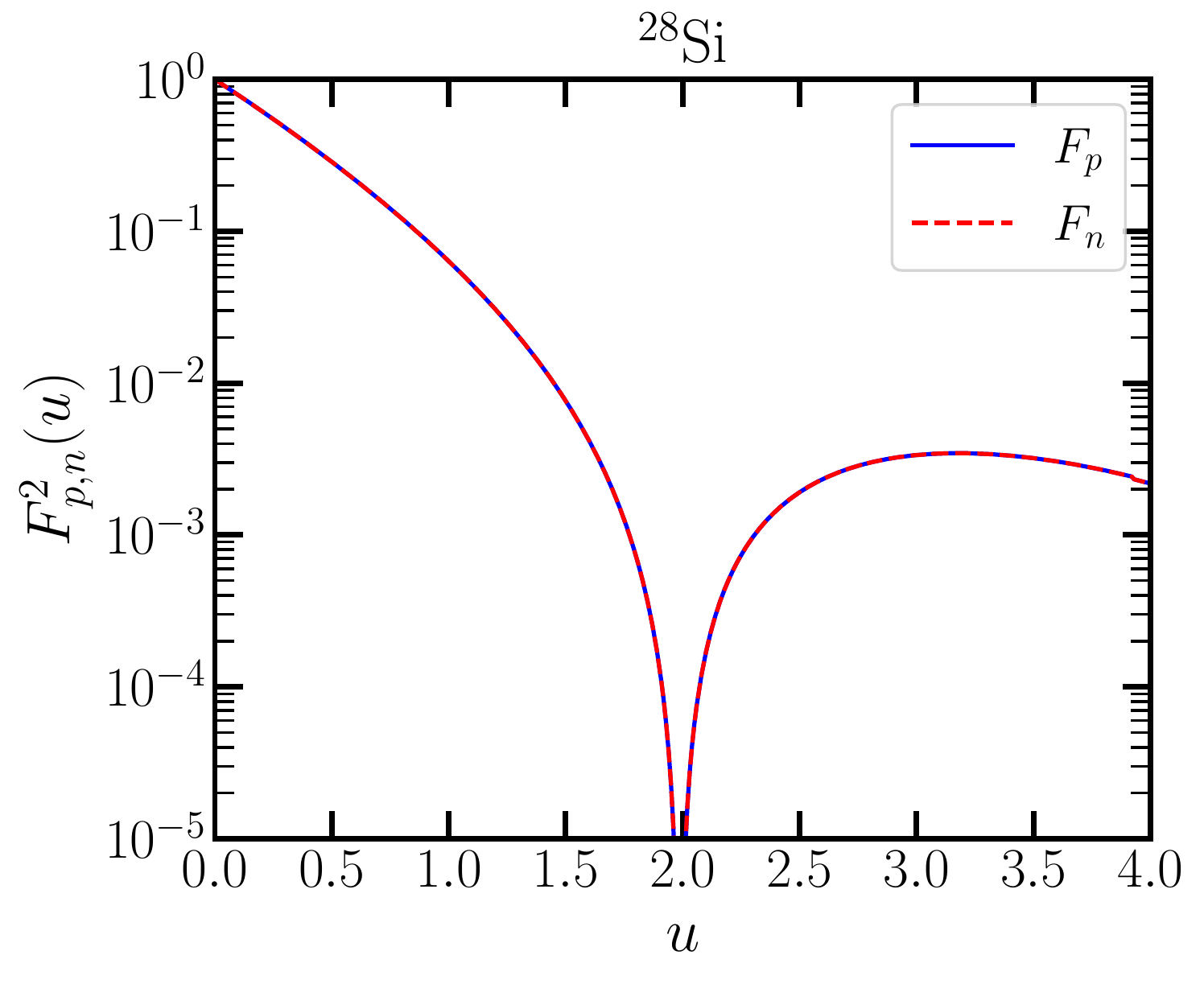}
	\caption{Square of proton (solid line)  and neutron (dotted line)
	form factors for $^{70,73,76}$Ge, 
	$^{64,70}$Zn and $^{28}$Si as a function of $u=q^2b_\ell^2/2$.}
	\label{fig:formfactor}
\end{figure*}

\subsection{Nuclear structure calculations within the DSM}

As is well known, the largest source of theoretical uncertainty in \cevns originates from the nuclear physics~\cite{Papoulias:2019lfi} (see also Ref.~\cite{Cadeddu:2017etk}). In this subsection, we evaluate the corresponding nuclear form factors with high reliability by incorporating our realistic nuclear structure DSM calculations. In Table~\ref{tab:1}, some useful nuclear structure properties of the isotopes studied in the 
present work are tabulated.
\begin{table*}[ht!]
\begin{tabular}{|l|ccc|cc|c|}
\hline
      \diagbox{Property}{Nucleus}          & \multicolumn{3}{c|}{Germanium}       & \multicolumn{2}{c|}{Zinc}     & Silicon       \\
      \hline
     & $^{70}$Ge  & $^{73}$Ge  & $^{76}$Ge & $^{64}$ Zn  & $^{70}$ Zn  & $^{28}$ Si  \\ 
\hline
ground state spin ($J^\pi$) & $0^+$             & $9/2^+$          & $0^+$            & $0^+$             & $0^+$            & $0^+$    \\ 
\hline
isotopic abundance (\%)     &  20.52 &  7.76  &  7.75  &  49.2  &  0.6  &  92.2  \\ 
\hline
h.o. length (fm)            & 1.894             & 1.907            & 1.920            & 1.865             & 1.894            & 1.625    \\ 
\hline
\end{tabular}
\caption{Nuclear structure properties of the studied isotopes (for details regarding the model space chosen, the sp energies, etc., 
see the text).}
\label{tab:1}
\end{table*}

In the following paragraphs, for the benefit of the reader we describe briefly some basic properties of the studied CE$\nu$NS 
detector materials.

\subsubsection{The studied isotopes as CE$\nu$NS detectors}

(1) For the $^{70}$Ge isotope, in our earlier double-$\beta$ decay (DBD) study using DSM~\cite{Sahu:2014nga}, calculations have been carried out for spectroscopic properties such as the energy spectra (band structures), the B(E2) values and occupancies.
For this isotope, the $jj$44b 
effective interaction in a model space consisted of the $^2p_{3/2}$, $^1f_{5/2}$, $^2p_{1/2}$, $^1g_{9/2}$ $j$-levels with 
single particle (sp) energies ($-9.6566$, $-9.2859$, $-8.2695$,$-5.8944$) MeV respectively~\cite{Cheal:2010zza}, gave good agreement with data.
Additional isotopic properties are listed in Table \ref{tab:1}. We note that for the harmonic 
oscillator (h.o.) size (length) parameter we used the simple expression $b_\ell = 0.933 \, A^{1/6} = 1.894$~fm for all the
nuclei studied.

(2)  For the $^{73}$Ge, the DSM calculations have been assessed through the spectroscopic properties in Ref.~\cite{Sahu:2017czz}
by employing the modified Kuo effective interaction~\cite{ABS} in the model space: $^2p_{3/2}$, $^1f_{5/2}$, $^2p_{1/2}$, $^1g_{9/2}$,
with the sp energies ($0$, $0.78$, $1.08$, $4.9$) MeV, respectively. Band structures and other spectroscopic properties are well reproduced 
with the aforementioned ingredients. We also mention that, in our recent CE$\nu$NS results for $^{73}$Ge in Ref.~\cite{Sahu:2020kwh} we
have adopted the same model parameter values with $b_\ell = 1.907$~fm. 

(3)  For the $^{76}$Ge isotope, which is a well-known neutrinoless double-$\beta$ decay (DBD) detector material, the DSM calculations 
have been performed with the modified Kuo interaction and the same model space as well as sp energies as in the case of the $^{73}$Ge 
isotope above ($b_\ell =1.920$ fm)~\cite{Sahu:2013yna}. Low-lying bands, $B(E2)$ values and orbit occupancies are well 
reproduced for this isotope too.

(4)  The DSM calculations for the $^{64}$Zn isotope have been carried out using GXPF1A effective interaction~\cite{Sahu:2013yna}.
The model space consisted of the ($^1f_{7/2}$, $^2p_{3/2}$, $^1f_{5/2}$, $^2p_{1/2}$) orbits with sp energies ($-8.6240$, $-5.6793$, 
$-1.3829$, $-4.1370$) MeV, respectively~\cite{Honma:2004xk}. The energy spectra, the $B(E2)$ values and occupancies of sp orbits (with
$b_\ell = 1.865$ fm) agree very well with the corresponding experimental data~\cite{Sahu:2013yna}.

(5) For $^{70}$Zn,  the DSM spectroscopic results  have been obtained with the ingredients of $^{70}$Ge as described above 
($b_\ell = 1.894$)~\cite{Sahu:2014nga}.

(6) For the $^{28}$Si, the spectroscopic calculations within DSM have been performed ($b_\ell = 1.625$ fm) by using 
the recently determined USD effective interaction~\cite{Brown:2006gx,Magilligan:2020bbd}. The calculated energy spectra, 
the $B(E2)$'s and also the $B(M1)$ values agree reasonably well with the experimental data and they will be discussed elsewhere 
\cite{Kosmas-in-preparation}.

Although we have used a formula for  $b_\ell$ following the DSM study of $^{72}$Ge in Ref.~\cite{Kosmas:2003xr}, it is desirable to deduce the values of $b_\ell$ for each isotope from proton
charge radii obtained
by electron scattering experiments. The experimental values for the charge radii
(in fm) are 4.041, 4.063, 4.081, 3.928, 3.985
and 3.122 for $^{70,73,76}$Ge, $^{64,70}$Zn and $^{28}$Si as given in Ref.~\cite{ANGELI201369}.
However, theoretical calculations for charge radii (see Ref.~\cite{BROWN1983313} for shell model and Ref.~\cite{Mukherjee:1984zz} for HFB examples) involve not only $b_\ell$ as a parameter but also
effective charges.
A consistent analysis of experimental data for charge form factors, charge radii,
quadrupole moments and $B(E2)$ values using DSM for the nuclei studied
here, will be considered in a future work. Finally, we note that 
the length parameter $b_\ell=0.933\, A^{1/6}$  (in fm) used (see Table~\ref{tab:1}) is slightly
smaller than the conventional parametrization $\hbar\omega=41 \, A^{1/3}$ giving
$b_\ell=0.994 \, A^{1/6}$.

\subsection{DSM calculations of the nuclear form factors needed for CE$\nu$NS event rates}

The proton and neutron nuclear form factors, $F_{p,n}(u)$, in terms of the dimensionless parameter $u=q^2(b_\ell)^2/2$, are illustrated in Fig.~\ref{fig:formfactor}. As can be seen from this figure, for the even-even 
isotopes $^{76}$Ge, $^{70}$Ge and $^{70}$Zn, the neutron form factor peaks shift toward smaller values of $u$. Again, the second 
neutron peak is larger than the corresponding proton peak. For the other three nuclear isotopes, the neutron and proton form factors 
are almost similar. Further details for the nuclei of interest, on the chosen nuclear configuration, the effective interaction, the 
$b_\ell$ value, etc., are listed in Table \ref{tab:1} and are discussed previously.

\begin{figure*}[ht!]
\includegraphics[width=0.4\linewidth]{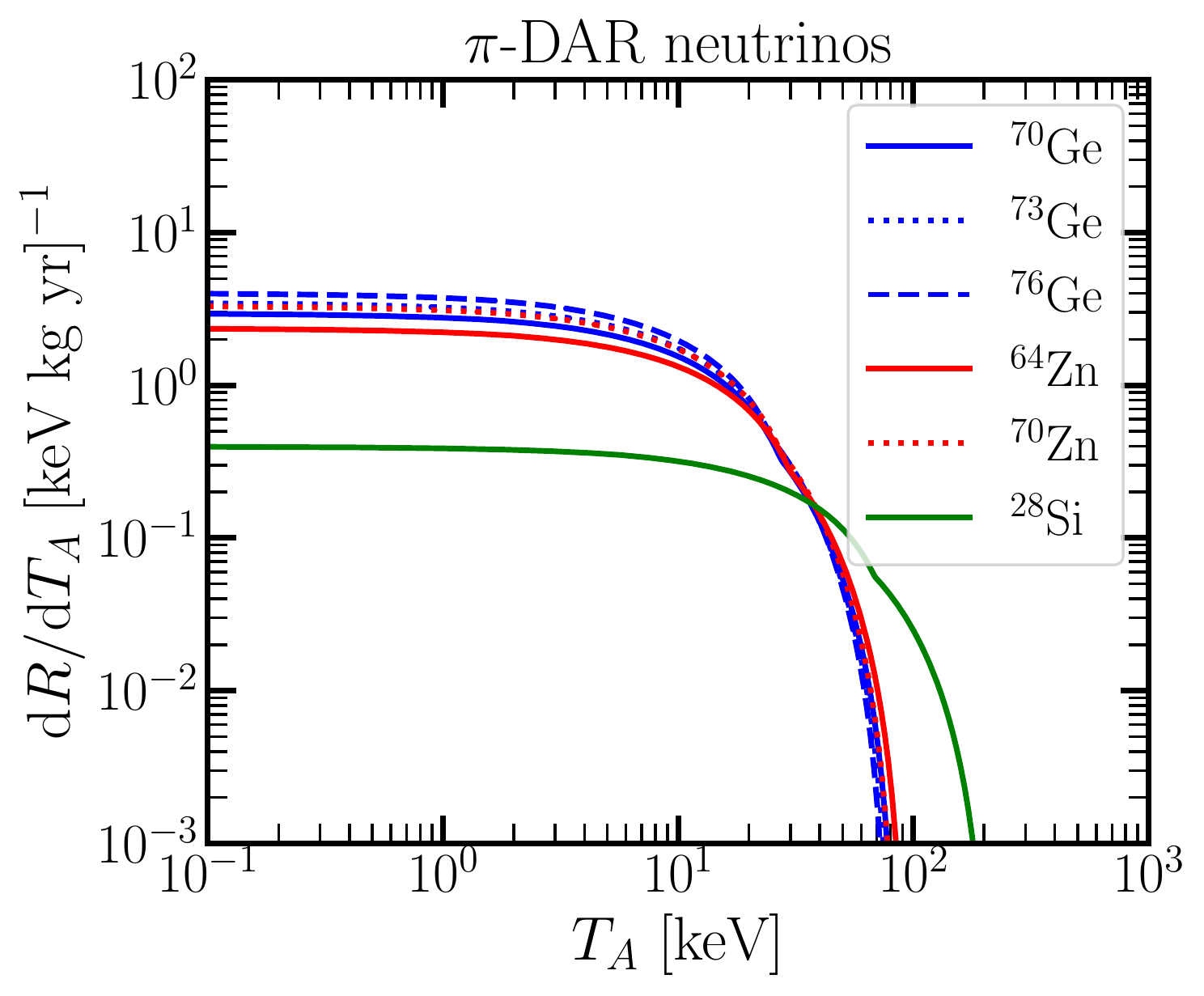}
\includegraphics[width=0.4\linewidth]{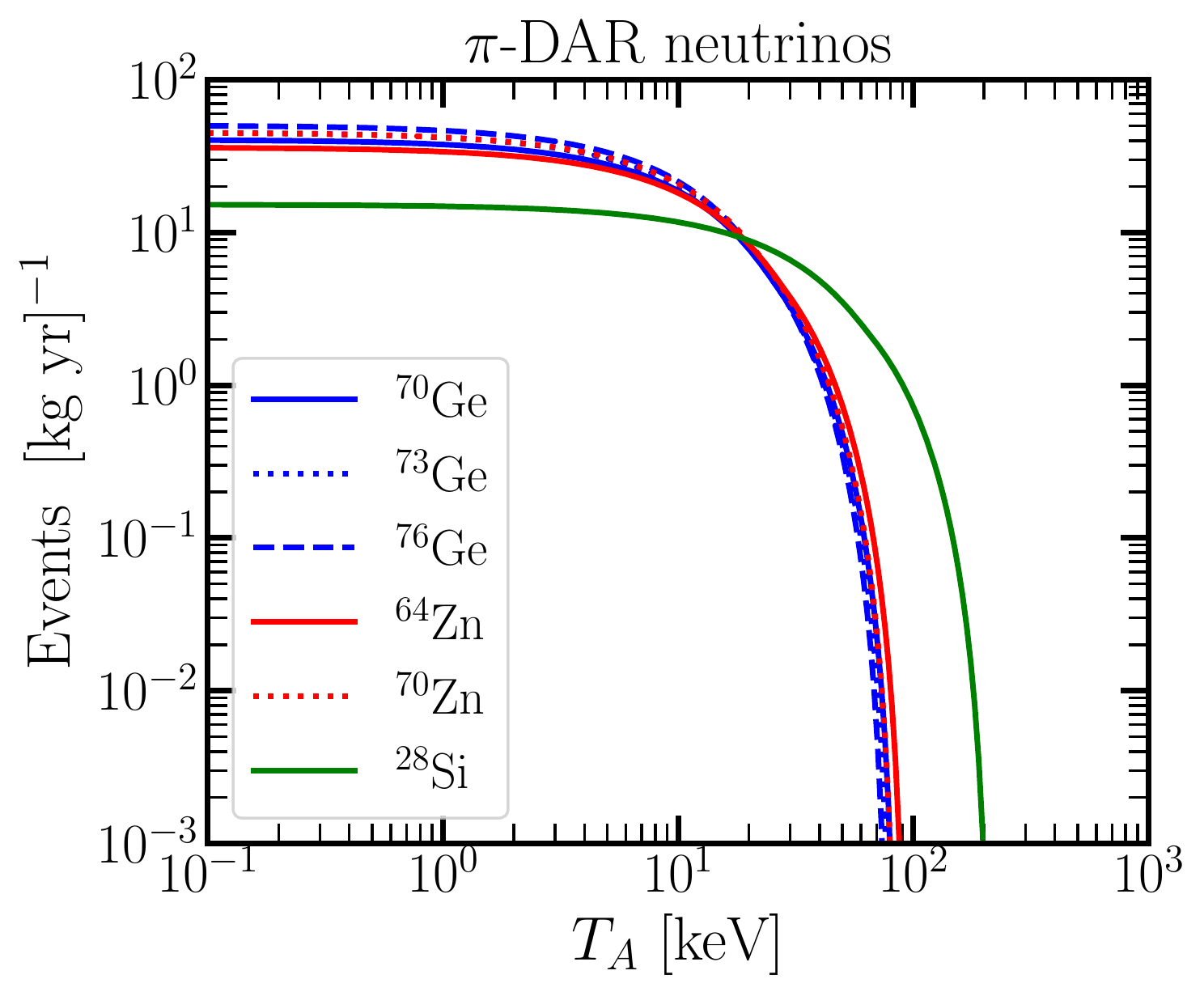}
	\caption{Differential (left) and integrated (right) event rates
	as a function of the nuclear recoil energy for $^{70,73,76}$Ge, $^{64,70}$Zn and $^{28}$Si.
The results are presented for CE$\nu$NS process with $\pi$-DAR neutrinos.}
\label{fig:sns}
\end{figure*}

\begin{figure*}[ht!]
\includegraphics[width=0.4\linewidth]{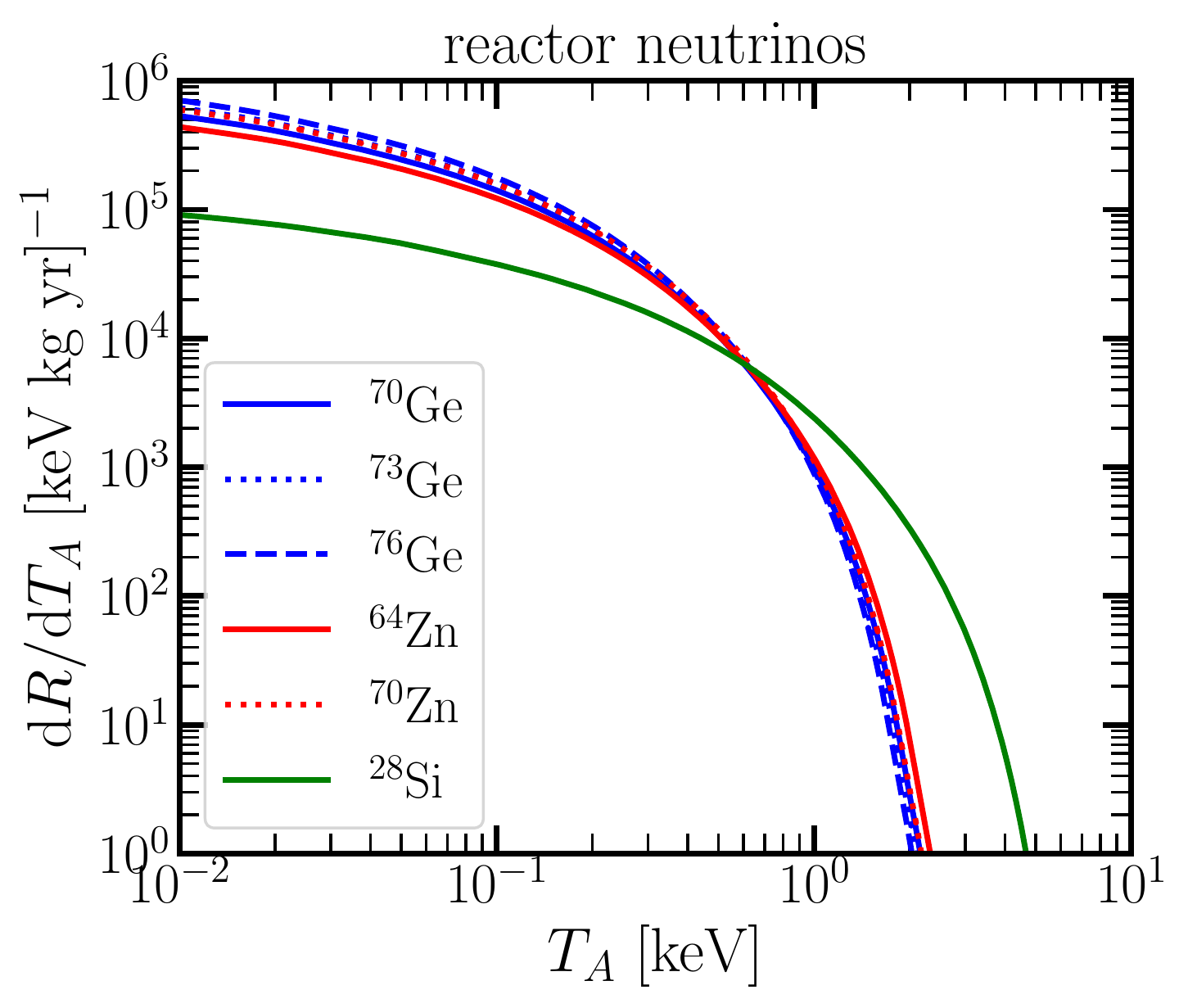}
\includegraphics[width=0.4\linewidth]{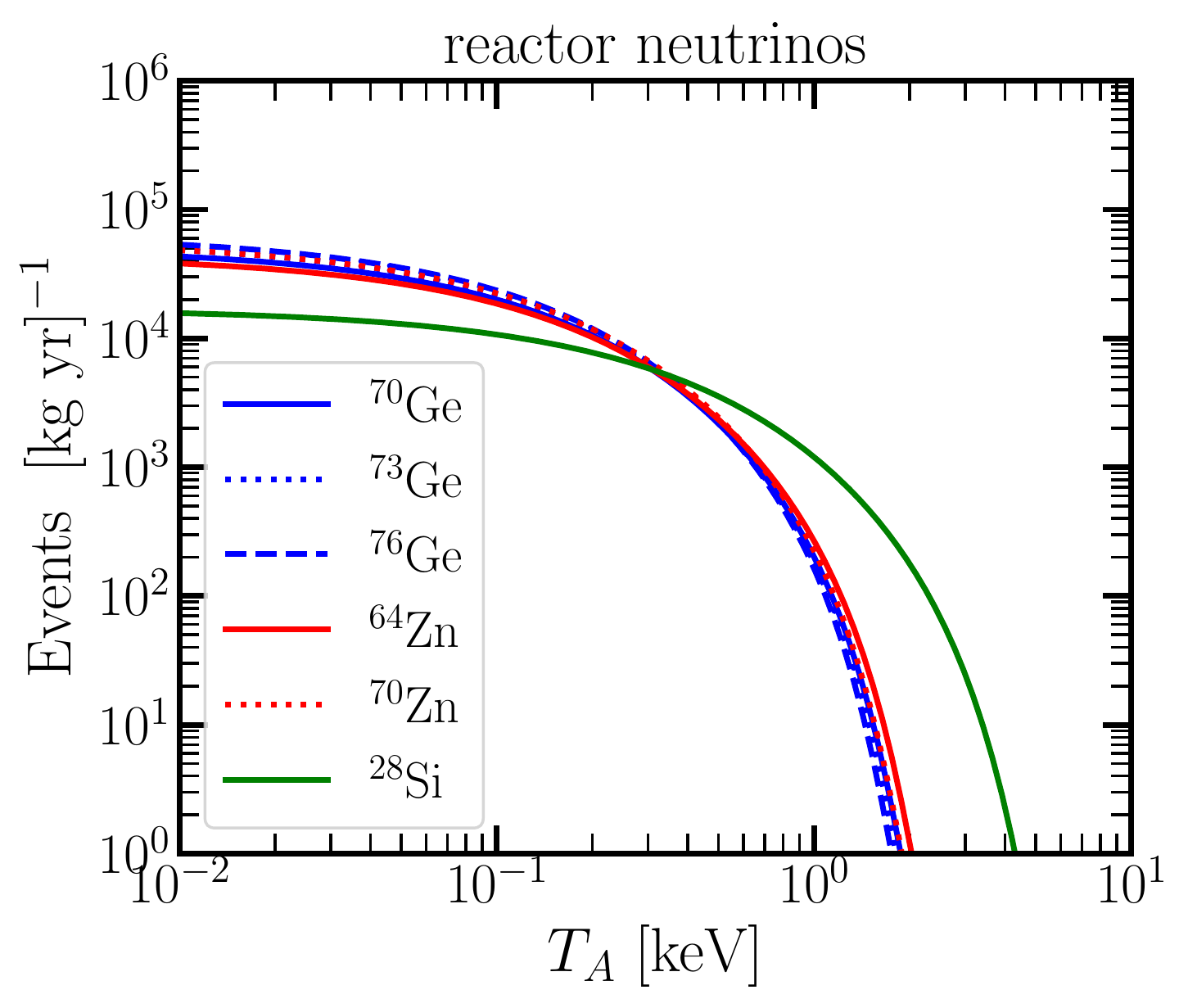}
\caption{Differential (left) and integrated (right) event rates
        as a function of the nuclear recoil energy for $^{70,73,76}$Ge,  $^{64,70}$Zn and $^{28}$Si.
The results are presented for CE$\nu$NS process with reactor neutrinos.}
\label{fig:reactor}
\end{figure*}

\begin{figure*}[ht!]
\includegraphics[width=0.4\linewidth]{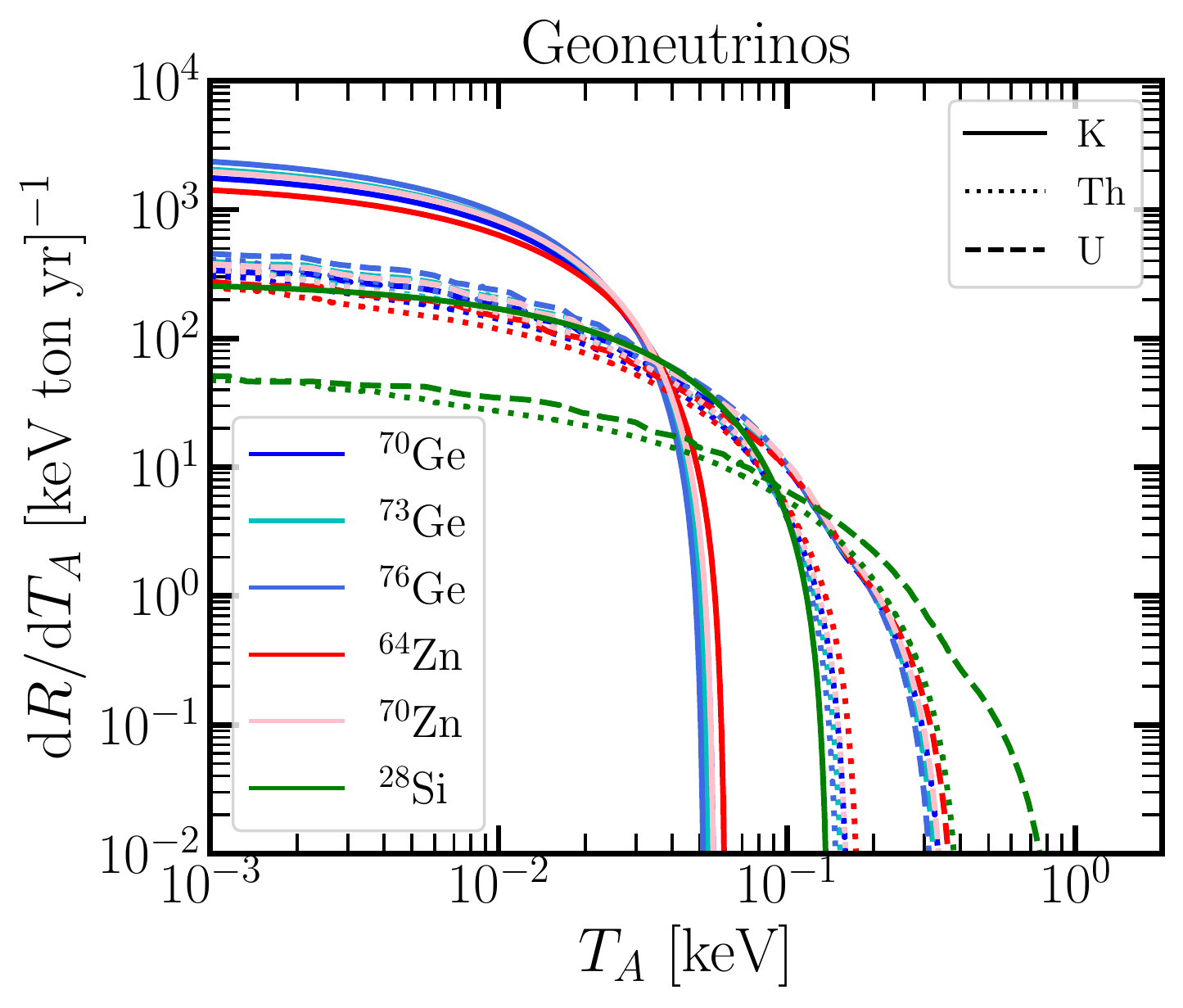}
\includegraphics[width=0.4\linewidth]{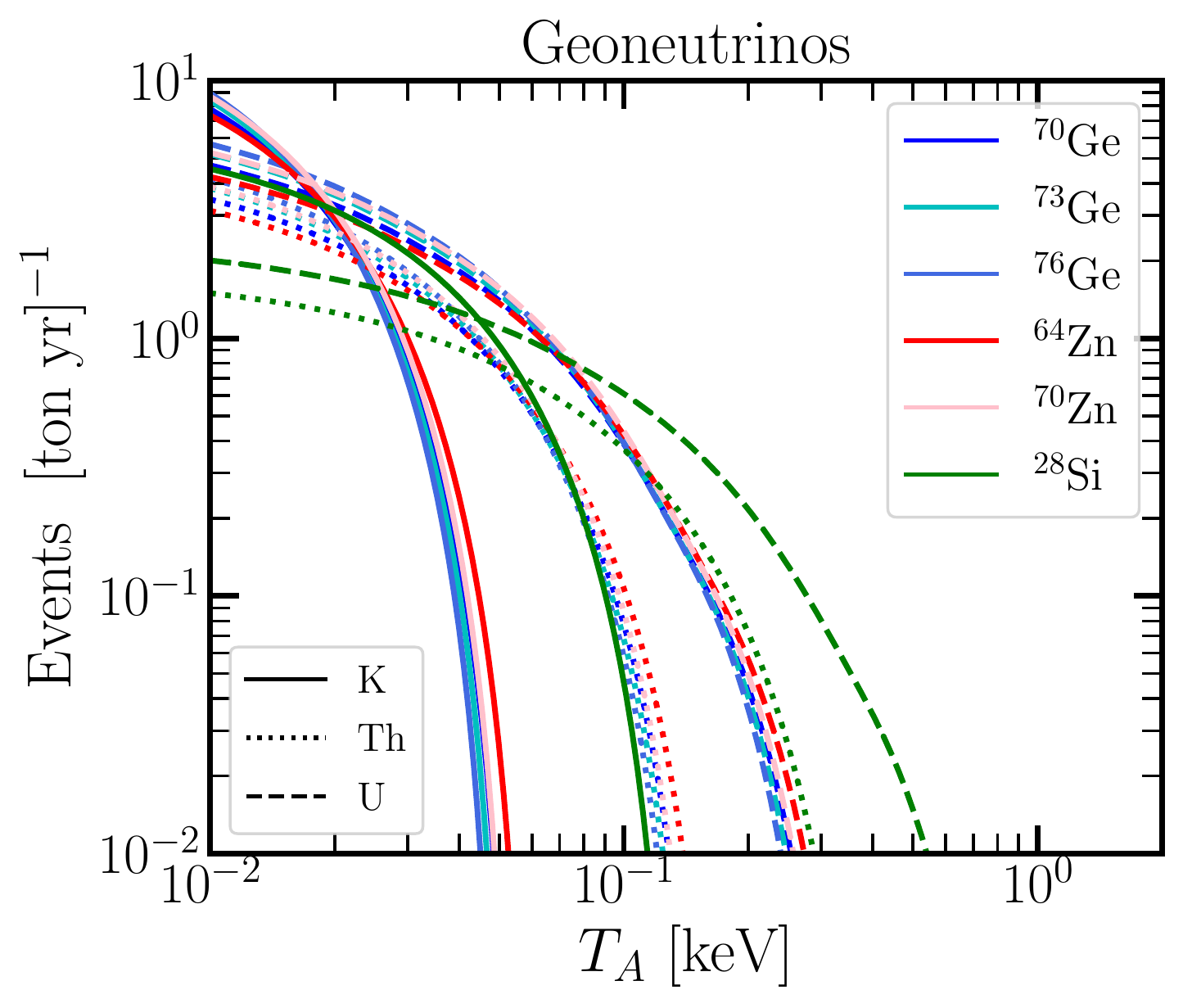}
\caption{Differential (left) and integrated (right) event rates
        as a function of the nuclear recoil energy for $^{70,73,76}$Ge,  $^{64,70}$Zn and $^{28}$Si.
The results are presented for CE$\nu$NS process with reactor neutrinos.}
\label{fig:geonu}
\end{figure*}

\begin{figure*}[ht!]
\includegraphics[width=0.4\linewidth]{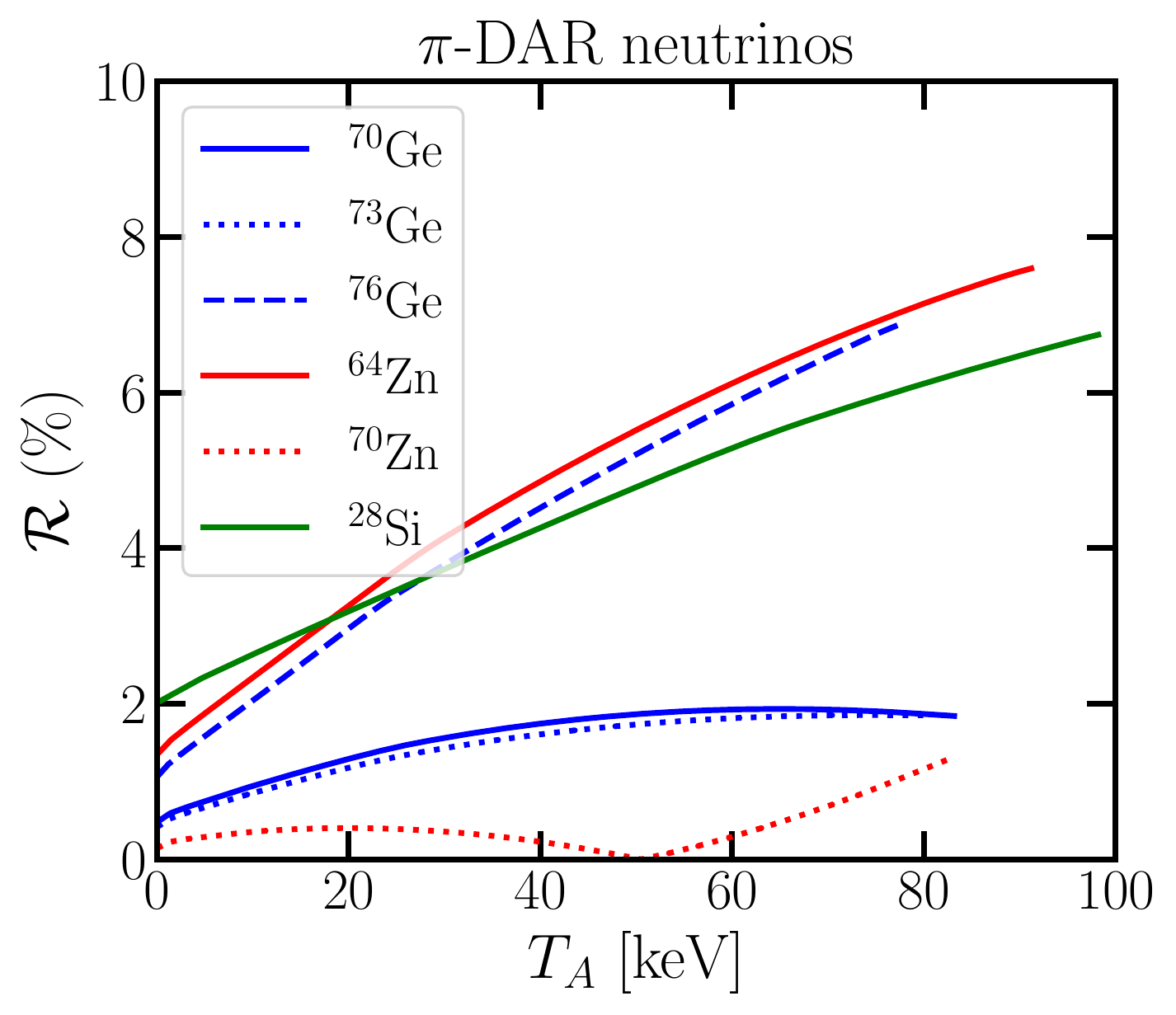}
\includegraphics[width=0.4\linewidth]{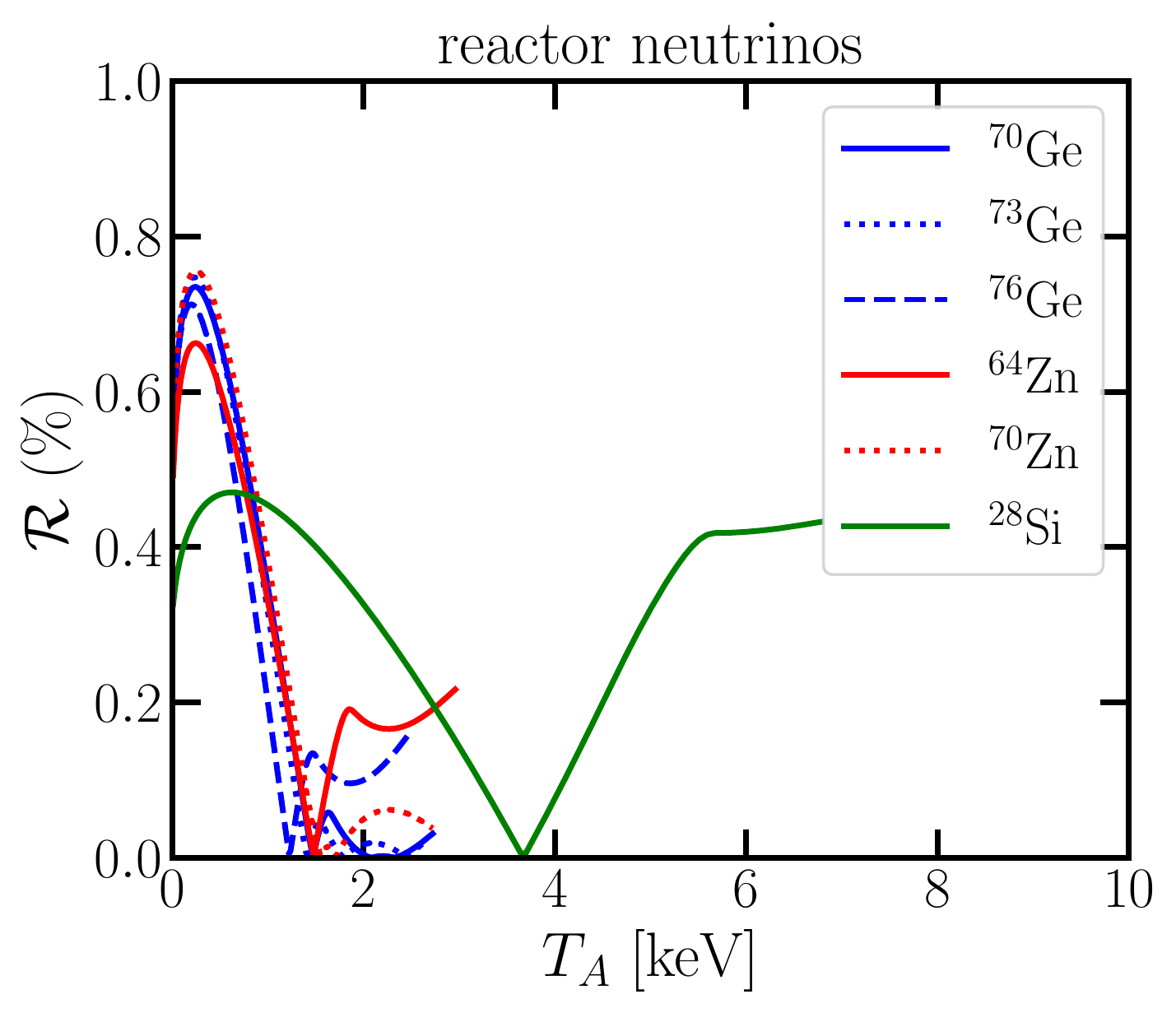}
\caption{Percentage difference between DSM calculations and those involving the effective Klein-Nystrand form factor parametrization 
(for details see the text). The results are presented for the case of $\pi$-DAR neutrinos (left) and reactor neutrinos (right)
        as a function of the nuclear recoil energy for $^{70,73,76}$Ge, $^{64,70}$Zn and $^{28}$Si.}
\label{fig:differences}
\end{figure*}

By employing the form factors shown in Fig.~\ref{fig:formfactor}, the differential and integrated event rates are
calculated utilizing Eq.(\ref{eq.6}) for neutrinos of the SNS, reactor and geoneutrino sources and the results are illustrated
in Figs. \ref{fig:sns}, \ref{fig:reactor} and \ref{fig:geonu}. In these plots, we have used different
isotopes of the same element which is crucial for reducing the systematic errors as pointed out in
Ref.~\cite{Galindo-Uribarri:2020huw}. For example, for SNS the proportion of differential event rates 
for $^{70, 73, 76}$Ge at recoil energy 0.1 keV is 2.96:3.64:4.0 which is equivalent to 1:1.230:1.351. 

Assuming that the $N^2$ dependence is approximately valid, the corresponding proportions are $38^2$:$41^2$:$44^2$, which is 
equivalent to 1:1.164:1.340. For the case of reactor neutrons, the corresponding proportions are 1:1.213:1.316. For other 
cases also, we obtain similar results. Use of detectors made up of a set of isotopes of an element may help in precision 
measurements at different experimental facilities. 

Turning to large-scale dark matter direct detection detectors, our current results indicate sizable geoneutrino-induced event 
rates, especially for sub-keV thresholds. Even though the detectable geoneutrino background signal will be completely dominated 
by solar neutrino events, it is expected to become a crucial component in the overall neutrino background at future ton-scale 
detectors looking for weakly interacting massive particles (WIMPs), especially for those aiming to detect low mass WIMPs with $m_\chi \leq 10 ~\mathrm{GeV/c^2}$. 
As a concrete example, we discuss the SuperCDMS experiment at SNOLAB which aims to reach nuclear recoil thresholds as low as 
40 eV (78 eV) using a germanium (silicon) detector~\cite{SuperCDMS:2016wui} for which our present calculations are particularly 
relevant and of significant importance.

Finally, we are interested to quantify the percentage difference on the number of events calculated using our nuclear structure 
DSM calculations or involving effective form factor approximations.  As a benchmark test case, we consider the Klein-Nystrand 
form factor approximation~\cite{Klein:1999qj} that has been recently adopted by the COHERENT Collaboration~\cite{Akimov:2017ade}. 
We illustrate the difference between the two calculations by evaluating the quantity
\begin{equation}
\mathcal{R}= \frac{|R_\text{DSM} - R_\text{KN}|}{R_\text{DSM}}
\end{equation}
and our corresponding results are shown in Fig.~\ref{fig:differences}. As can be seen, reactor neutrino experiments looking for 
\cevns will not suffer from nuclear structure uncertainties, even at the sub-percentage level.  On the other hand, for the case 
of $\pi$-DAR neutrinos which involve larger values of the momentum transfer, $\mathcal{R}$ can be as high as 8\% for $^{64}$Zn 
and $^{76}$Ge. We finally note that here we do not present the corresponding results for geoneutrinos since the signal uncertainty 
will be dominated by the flux uncertainties, while also the momentum transfer is lower compared to reactor neutrinos. For solar, 
diffused supernova background and atmospheric neutrinos, such results have been presented in a previous study~\cite{Papoulias:2018uzy}.

\subsection{Application to nonstandard interactions}

\begin{figure*}[ht!]
\includegraphics[width=0.4\linewidth]{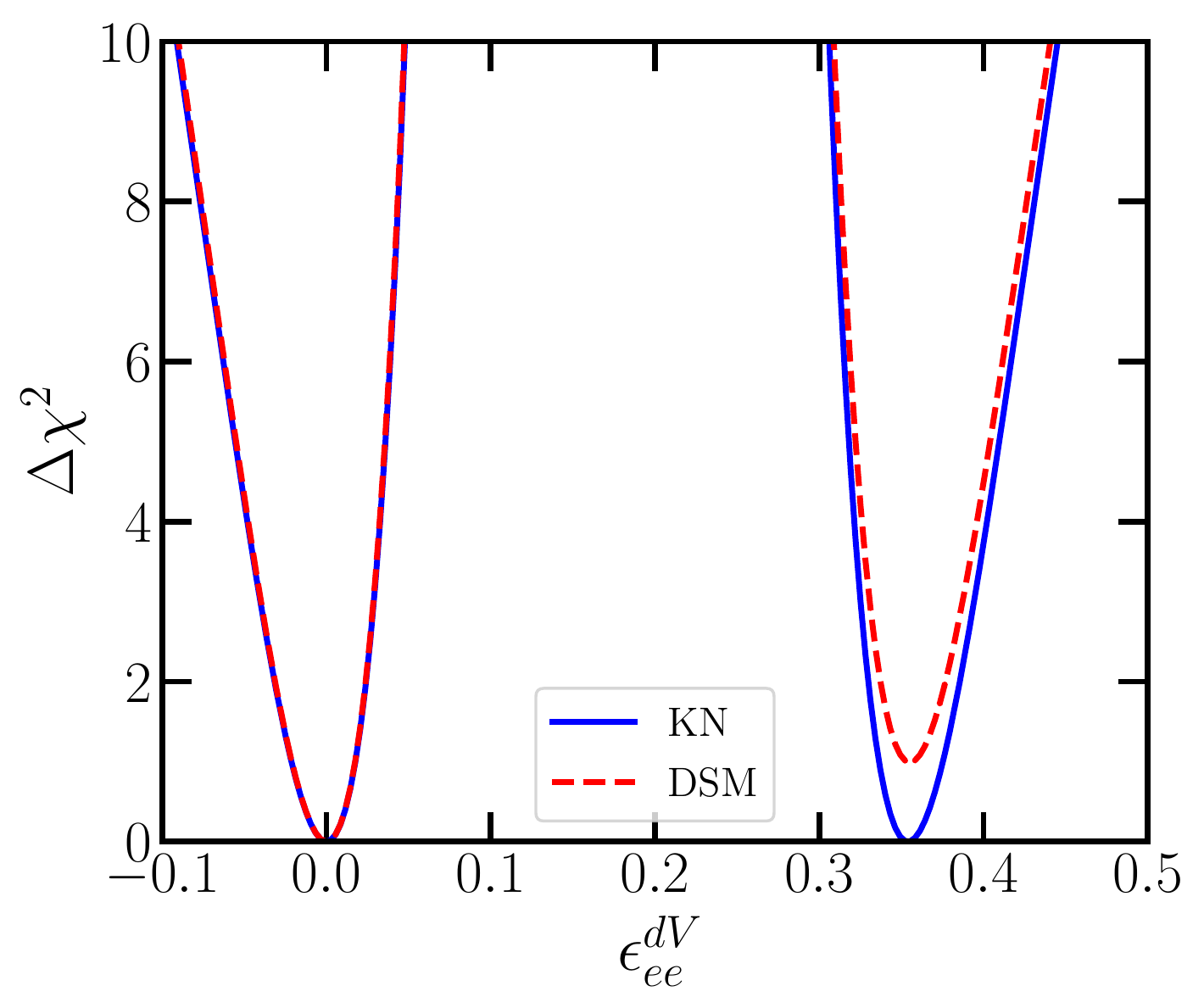}
\includegraphics[width=0.4\linewidth]{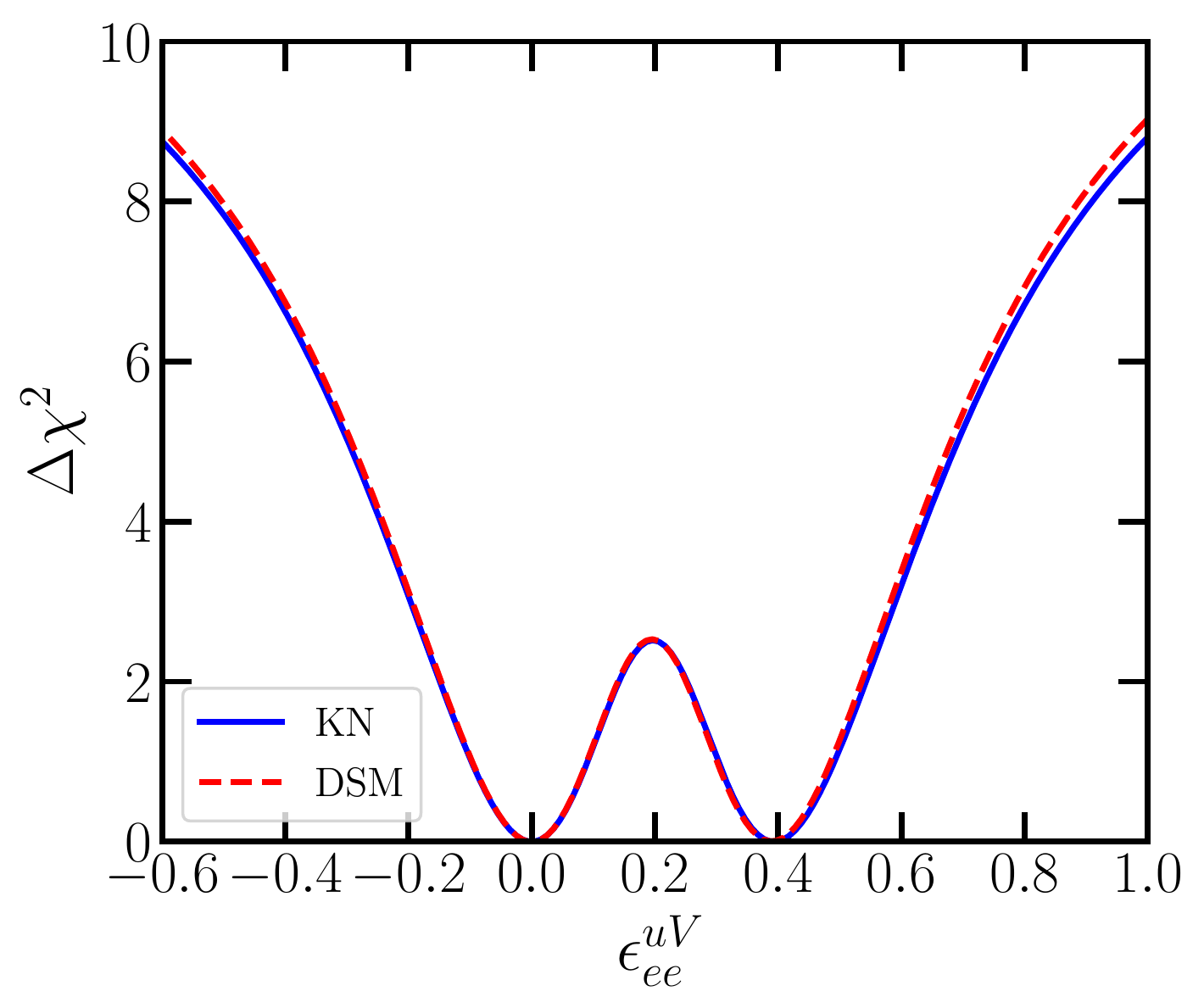}
\caption{Projected sensitivity to NSIs for a 10-kg $^{76}$Ge detector at a $\pi$-DAR facility. A comparison of the expected sensitivities is given assuming DSM and KN nuclear form factors.}
\label{fig:NSI}
\end{figure*}

Focusing on vector-type NSIs only, our goal is to explore the impact of DSM form factors to the projected NSI sensitivities. In order to quantify the effect of nonzero NSI contributions to the \cevns cross section, it is sufficient to replace the SM weak charge of Eq.(\ref{eq:Qw}) with the corresponding NSI charge according to the substitution
$\mathcal{Q}_W \to \mathcal{Q}_\text{NSI}$, with 
\begin{widetext}
\begin{equation}
\begin{aligned}
\mathcal{Q}^V_\text{NSI} =& 
  \left [ \left ( g_{V}^{p}+2\varepsilon _{\alpha \alpha}^{uV}+\varepsilon _{\alpha \alpha}^{dV}  \right )Z F_{p}(Q^{2})+\left ( g_{V}^{n}+\varepsilon _{\alpha \alpha}^{uV}+2\varepsilon _{\alpha \alpha}^{dV}  \right ) N  F_{n}(Q^{2})\right ]  \nonumber \\
  +&\sum_{\alpha }\left [ \left ( 2\varepsilon _{\alpha \beta}^{uV}+\varepsilon _{\alpha \beta}^{dV} \right ) Z F_{p}(Q^{2})+ \left ( \varepsilon _{\alpha \beta}^{uV}+2\varepsilon _{\alpha \beta}^{dV} \right )N F_{n}(Q^{2})    \right ] .
  \end{aligned}
\label{eq:qvNSI}
\end{equation}
\end{widetext}
Our sensitivity analysis is based on a simple $\chi^2$ function
\begin{equation}
\chi^2 = \sum_{i=1}^{50} \left( \frac{R_\text{SM}^i - (1+a)R_\text{NSI}^i(\epsilon_{ee}^{uV}, \epsilon_{ee}^{dV})}{\sigma_\text{stat}^i} \right)^2 + \left( \frac{a}{\sigma_a}\right)^2 \, ,
\end{equation}
for which we consider 50 equal-size bins of recoil energy in the range 5--80 keV, allowing for non-zero NSIs with the $\nu_e$ flux only. The statistical uncertainty is defined as $\sigma^i_\text{stat}= \sqrt{R^i_\text{SM} +  R^i_\text{bkg}}$, assuming a flat background $R^i_\text{bkg} = \sigma_\text{bkg} R^i_\text{SM}$. We furthermore consider a conservative scenario taking the background and signal uncertainties to be $\sigma_\text{bkg}=\sigma_a =30\%$.  For our statistical analysis, we assume a $\pi$-DAR neutrino source with a 10-kg $^{76}$Ge target nucleus for which we expect the impact of nuclear form factors to be maximized (see left panel of Fig.~\ref{fig:differences}).

Taking one nonvanishing NSI parameter at a time, our results are presented in Fig.~\ref{fig:NSI}. As can be seen from the $\chi^2$ profiles, the use of DSM or KN form factors will not alter the sensitivities on $\epsilon_{ee}^{uV}$. On the other hand, for the case of $\epsilon_{ee}^{dV}$  our fit clearly prefers the trivial solution over the nonzero one when relying on DSM nuclear structure calculations. This is found to be in contrast to the case of KN calculations where there is absence of a best-fit point preference. A few comments are in order. As expected, the sensitivity on $\epsilon_{ee}^{dV}$ is stronger compared to $\epsilon_{ee}^{uV}$, and hence the implications of DSM calculations are more pronounced in the former case, which might be helpful for resolving the LMA dark degeneracy~\cite{Coloma:2017egw}. As a final remark, we have checked that the neutrino-floor explored in Ref.~\cite{AristizabalSierra:2021kht}, due to the currently large uncertainties, is not affected by the choice of phenomenological or nuclear structure form factors.

\section{Conclusions}
\label{sec:conclusions}

Our main aim in the present study was to perform calculations of the \cevns event rates for the Ge-detectors chosen in ongoing
and designed \cevns experiments. We also studied Zn and Si which are considered promising target materials of experiments aiming 
to measure \cevns events. The nuclear structure calculations have been carried out (for the specific isotopes $^{70,73,76}$Ge, 
$^{64,70}$Zn and $^{28}$Si) with a high level of reliability, by taking into account crucial information from the nuclear structure. 
The detailed nuclear physics aspects came out of the DSM method which involves realistic two-body interactions and is assessed on 
the reproducibility of experimental microscopic nuclear properties. 

Highly accurate calculations such as those provided here, are valuable for discriminating the expected signal from the various isotopic 
admixtures contained in germanium or zinc detectors, the use of which has been proposed for reducing the experimental uncertainties. 
We have considered typical experimental configurations, exposed to neutrinos from $\pi$-DAR, reactor antineutrinos and geoneutrinos, 
while to the best of our knowledge, the present work is the first nuclear-physics-based study with regards to geoneutrino signals. 

We compared our theoretical event rates with those calculated on the basis of the widely adopted form factors (e.g., the 
phenomenological Klein-Nystrand) and we concluded that especially for the SNS neutrinos the differences can be of the order of 10\%. 
On the other hand, we have verified that reactor antineutrino facilities with sub-keV thresholds as well as large-scale direct dark matter 
detection experiments looking for light WIMPs uncertainties may be neglected for very low momentum transfer involved in the \cevns process. We have finally discussed the robustness of the attainable sensitivities on NSI with regards to phenomenological and DSM nuclear form factors.

\acknowledgments
The research of D.K.P. is co-financed by Greece and the European Union (European Social Fund- ESF) through the Operational Programme «Human Resources Development, Education and Lifelong Learning» in the context of the project ``Reinforcement of Postdoctoral Researchers - 2nd Cycle'' (MIS-5033021), implemented by the State Scholarships Foundation (IKY).
R.S. is thankful to SERB of Department of Science and Technology (Government of
India) for financial support. The work of T.S.K. is implemented through the Operational Program ``Human Resources Development,
Education and Lifelong Learning - cycle B''  (MIS-5047635) and is co-financed by the European Union (European Social Fund) and Greek national funds.


\providecommand{\href}[2]{#2}\begingroup\raggedright\endgroup

\end{document}